\documentclass{JHEP3}

\input{epsf.sty}
\usepackage{epsfig,amssymb, amsmath, graphicx}

\def\be{\begin{equation}}
\def\ee{\end{equation}}    
\def\ba{\begin{eqnarray}}
\def\ea{\end{eqnarray}}

\def\2pi{\left(2\pi\right)}

\def\df{\delta\phi}

\newcommand{\Expect}[1]{\left\langle #1 \right\rangle}

\title{The structure of correlation functions in single field inflation}
\author{Sarah Shandera
\footnote{sarah@phys.columbia.edu}
 \\ Institute of Strings, Cosmology and Astroparticle Physics,
Physics Department, Columbia University, New York, NY 10027}

\abstract{
Many statistics available to constrain non-Gaussianity from inflation are simplest to use under the assumption that the curvature correlation functions are hierarchical. That is, if the $n$-point function is proportional to the $(n-1)$ power of the two-point function amplitude and the fluctuations are small, the probability distribution can be approximated by expanding around a Gaussian in moments. However, single-field inflation with higher derivative interactions has a second small number, the sound speed, that appears in the problem when non-Gaussianity is significant and changes the scaling of correlation functions. Here we examine the structure of correlation functions in the {\it most general} single scalar field action with higher derivatives, formalizing the conditions under which the fluctuations can be expanded around a Gaussian distribution. We comment about the special case of the Dirac-Born-Infeld action.}

\begin{document}

\maketitle

\section{Introduction}
The next decade will bring extraordinarily detailed measurements of the cosmic microwave background from the Planck satellite \cite{Planck}. In addition, large volume surveys will catalogue cosmological structures (such as galaxies, galaxy clusters, and voids) whose number and distribution depends on the same initial conditions that explain the temperature fluctuations\footnote{Relevant current and upcoming surveys include SDSS, BOSS, WFMOS, ADEPT, EUCLID, DES, PanSTARRS, LSST.}. This new data will enable an increasingly precise determination of the temperature, polarization and matter power spectra over a widening range of scales. More importantly, it will also allow measurements of higher order statistics. These statistics are an excellent probe of inflation, in theory measuring interactions of the inflaton that can distinguish qualitatively different inflationary physics and perhaps truly testing inflation itself. 

If the connected part of any primordial higher order correlation function is observed to be different from zero, then in the inflationary paradigm the inflaton was not a free field and its fluctuations (which source the observed temperature and matter density contrasts) were non-Gaussian. Clearly, at least a small deviation from Gaussianity is expected. Furthermore, models with derivative self-interactions or multiple fields (including the alternative ekpyrotic scenario) can generate significant non-Gaussianity \cite{Seery:2005wm, Chen:2006nt, Bartolo:2001cw, Bernardeau:2002jy, Bernardeau:2002jf, Buchbinder:2007at, Lehners:2007wc}. This is in contrast to standard single-field slow-roll with a smooth potential, where the level of non-Gaussianity is proportional to slow-roll parameters \cite{Acquaviva:2002ud, Maldacena:2002vr}. Such non-Gaussianity is so small that its detection must likely wait, even in the most optimistic analysis, at least until measurements of the 21cm line \cite{Cooray:2006km} and even so must be separated from gravity induced effects that are an order of magnitude or two larger \cite{Pyne:1995bs, Bartolo:2006cu, Bartolo:2006fj, Pitrou:2008ut}. Here we will focus on the single-field scenario where the primordial non-Gaussianity is mostly generated by derivative interactions and is observably large. This has the dual advantage of being observationally relevant now and calculationally simple. 

Much of the work on the observational implications of non-Gaussianity has so far focused on the CMB bispectrum (3-point function) for the excellent reason that this is likely to be a statistic where a deviation from Gaussianity is easy to detect. More generally, tools for evaluating non-Gaussianity come in two types: explicit comparison of correlation functions predicted by a model to the data (so far, the 3-point or bispectrum as described above) and interpretation of other measured statistics (such as cluster number counts \cite{Lucchin:1987yv, Colafrancesco:1989px, Chiu:1997xb, Robinson:1998dx, Robinson:1999wh, Koyama:1999fc, Matarrese:2000iz, Verde:2000vr} or Minkowski functionals \cite{Hikage:2006fe, Hikage:2008gy}). The first case is of course ideal if we are confident that we know the correct model. However, non-Gaussianity is an excellent tool for distinguishing between different inflationary physics, so it is worthwhile to check for non-Gaussian features of the data in a variety of ways. Statistics like Minkowski functionals capture information about more than one correlation function and must to be interpreted to give, for example, constraints on the magnitude of the 3-point function (e.g., $f_{NL}$ for the local ansatz). 

At the moment, the translation between measured statistics and inflationary theory depends largely on the hierarchical assumption that the $n$-point correlation scales like the two-point to the $n-1$ power \cite{Matsubara:2003yt}. This assumption means in practice that one can expand around the Gaussian result in moments of the true distribution, as long as the amplitude of fluctuations is also small. It is a property of the non-Gaussianity generated by gravity after horizon re-entry (see e.g. \cite{Bernardeau:2001qr} for a thorough discussion) and is implicit in the local ansatz where the parameter $f_{NL}$ originates. However, the conditions under which a fundamental physics model gives rise to hierarchical correlations are only beginning to be discussed. The three and four point correlations from terms in the potential were discussed in \cite{Bernardeau:2003nx}, and there was earlier interest in models with dimensional scaling, where the $n$-point is proportional to the 2-point to the $n/2$ power (\cite{Bernardeau:2001qr} and references therein). When there are small parameters other than the root-mean-square amplitude in the problem, the scaling of the $n$-point with the two-point no longer ensures that a Gaussian expansion can be made. Here we consider `hierarchical' to mean that the distribution of curvature fluctuations can be expanded in a series of moments about a Gaussian (in a way that is made explicit in Section \ref{PXphi}). As our ability to observe higher order statistics improves, it is important to be clear about the structure expected from various physical processes to avoid misinterpreting the data. 

In this paper, we will examine the conditions under which the hierarchical assumption is justified for single-field models, where a second small number, the sound speed, appears in the correlation functions. Intuitively, the hierarchical structure in a single field model looks like a consequence of perturbativity, and we will show that there is indeed a single constraint that ensures this structure and the validity of perturbative calculations. Following \cite{Creminelli:2003iq, Cheung:2007st, Weinberg:2008hq, Leblond:2008gg, ArmendarizPicon:2008yv}, we will take an effective field theory (EFT) approach. More specifically, we will consider the EFT for the scalar field sourcing inflation, rather than the theory of the fluctuations. We study the most general single-field action and extend the results of \cite{Weinberg:2008hq}. We do this with the goal of understanding the structure of correlation functions from derivative interactions, and hope that some useful features can be uncovered even if a complete `EFT for inflation' is not fully understood. Field theory for inflation is being treated with much more care in some recent work \cite{Weinberg:2005vy, Weinberg:2006ac, Weinberg:2008hq, Weinberg:2008mc, Weinberg:2008nf} and it will be very interesting to verify that there is no subtlety that might change our conclusions.

Specifically, we will consider a single field description of the canonical inflaton $\phi$, valid below some scale $M\lesssim M_p$ (the reduced Planck mass). We take the point of view that this field exists regardless of whether there is also some physics providing a potential suitable for inflation. That is, we use flat space effective field theory intuition to construct the derivative terms in the effective action, then consider supplementing it with an appropriate potential energy $V(\phi)=3M_p^2H^2$ (where $H$ is the Hubble parameter) and computing the resulting spectrum of scalar fluctuations. Since we will only be interested in examining the correlation functions in a regime where the dominant non-Gaussian contribution comes from kinetic terms, we will not worry about the details of $V(\phi)$. 

Discussions of single field inflation have so far largely focused on actions of the form
\be
S=\int d^4x \sqrt{-g} [P(X,\phi) +\frac{1}{2}M_p^{2}R]
\label{fluidact}
\ee
where $X=-\frac{1}{2}g^{\mu\nu}\partial_{\mu}\phi\partial_{\nu}\phi$ \cite{Armendariz-Picon:1999rj}. For inflationary purposes (where the background solution is spatially homogeneous) the scalar part of the action describes an ideal fluid. This can be seen by computing the stress-energy tensor or more elaborately from symmetry arguments \cite{Dubovsky:2005xd}. We will review the cosmology associated with this action in the next section, including the hierarchical structure of correlations as long as a perturbative condition is satisfied. However, the $P(X,\phi)$ action is not the most general effective action for a scalar field $\phi$ since it contains no higher derivative terms. At the level of four-derivative terms, it was recently shown that this makes no difference \cite{Weinberg:2008hq} but cases with large non-Gaussianity require an understanding of terms with more derivatives. We will examine the most generic action up to six-derivative terms and show that the same perturbative condition still maintains the hierarchical structure. Extending to more derivatives uncovers no new features in this regard. The Dirac-Born-Infeld (DBI) action is a particularly interesting example of a $P(X,\phi)$ action that contains a square-root summation of powers of $X$. It is the action of a D-brane in the limit of small acceleration (small curvature), with higher derivative corrections whose structure is known in many contexts and can be guessed for the case relevant here. Because of this special structure, we will return to this example after discussing the more generic case. 

Higher derivative corrections are of little concern for the background inflating solution, $H(\phi_0)$, which anyway requires $\ddot{\phi_0}\ll H\dot{\phi_0}$, but it is less clear that these terms can be ignored in the correlation functions when the extra derivatives act on the fluctuations $\delta\phi$. Examining six-derivative terms in a generic scalar field action will uncover the condition under which these terms do not contribute significantly to correlation functions. 

In the next section, we further motivate our interest in the structure of correlation functions and review some basic notation for general single field models and inflationary calculations. In Section \ref{expanding} we extend the analysis of \cite{Weinberg:2008hq} to include the most relevant six derivative terms. Section \ref{pert} discusses the general structure of higher derivative terms in the perturbative regime and contains the main result. We turn to the illustrative example of the DBI action in Section \ref{DBI}, and contrast the likely form of the action with the general formalism derived in \cite{Cheung:2007st}. Finally, Section \ref{conclude} summarizes and concludes. Appendix \ref{Honest} demonstrates that calculation of the correlation functions for the action expanded to four or six derivatives gives the same scaling arrived at in Section \ref{pert} in a quick but approximate way.

\section{Review of inflation from $P(X,\phi)$}
\label{PXphi}
Here we review our notation for the power spectrum and cumulants, and the hierarchical ansatz. We also demonstrate the scaling behavior of correlation functions from derivative interactions in an action whose scalar part is $P(X,\phi)$ (Eq.(\ref{fluidact})).
\subsection{Inflationary Fluctuations}
We consider a single-field inflationary scenario where the inflaton field is $\phi=\phi_0+\delta\phi$ (we will work largely in the spatially flat gauge). For the action in Eq.(\ref{fluidact}), the procedure for quantizing scalar fluctuations proceeds just as for a field with a standard kinetic term, up to the appearance of an additional parameter, the sound speed, defined by 
\be
c_s^2=\frac{P_{,X}}{P_{,X}+2XP_{,XX}}
\label{definecs}
\ee
where $P_{,X}$ is the partial derivative of $P$ with respect to $X$ \cite{Armendariz-Picon:1999rj, Garriga:1999vw, Mukhanov:2005sc, Bean:2008ga}. Scalar fluctuations on a scale $k$ freeze out when $aH=c_sk$, where $H=\dot{a}/a$ and $a$ is the scale factor. Inflation is a period of accelerated expansion, which occurs as long as $\epsilon<1$ for
\be
\epsilon\equiv-\frac{\dot{H}}{H^2}\;.
\ee
Fluctuations have mean zero and amplitude
\be
\label{phivariance}
\Expect{\df^2}^{1/2} = \frac{H}{2\pi \sqrt{c_s P_{,X}}}
\ee
which reduces to the familiar answer for a canonical kinetic term. Using the lowest order gauge transformation, $\zeta=-(H/\dot{\phi_0})\delta\phi$, we can go from the spatially flat gauge to the comoving gauge to find the amplitude of primordial curvature fluctuations
\ba
\Expect{\zeta^2}^{1/2} =\frac{H}{2\pi M_p\sqrt{\epsilon c_s}}\;.
\ea
More precisely, the two-point function is defined 
\ba
\Expect{\zeta(\vec{k}_1) \zeta(\vec{k}_2)} &\equiv& (2\pi)^3\delta^3(\vec{k}_1+\vec{k}_2) P_\zeta\\ \nonumber
&=&(2\pi)^3\delta^3(\vec{k}_1+\vec{k}_2)2\pi^2\mathcal{P}_{\zeta}k^{-3}\\\nonumber
&=&(2\pi)^3\delta^3(\vec{k}_1+\vec{k}_2)\left(\frac{H^2}{4\pi^2M_p^2\epsilon c_s}\right)k^{-3}
\ea
where $\mathcal{P}_{\zeta}$ is the dimensionless power spectrum (variance). 

\subsection{Some notation for cumulants}
The power spectrum, or two-point function, is all that is needed to characterize a Gaussian field. Here we are interested in non-Gaussianity, which is a catch-all term indicating that additional information is needed to characterize the fluctuations. To structure the discussion, we review some notation for general probability distributions. We use the example of the Probability Distribution Function (PDF) for density fluctuations, $\delta$, to review how an observable may depend on the structure of correlation functions. This procedure can be generalized to apply to other quantities and other statistics, including the Minkowski functionals \cite{Matsubara:2003yt}.

For a probability density function $P(\delta)d\delta$, the $n$-th central moment is
\be
\langle\delta^n\rangle\equiv\int_{-\infty}^{\infty}\delta^nP(\delta)d\delta\;.
\ee
The $n$-th cumulant is the connected $n$-point function and cumulants with $n\geq3$ are zero for a Gaussian distribution. The reduced (or normalized) cumulants are defined as 
\be
\mathcal{S}_n\equiv\frac{\langle\delta^n\rangle_c}{\langle\delta^2\rangle_c^{n-1}}.
\ee
Then an exact expression for the PDF in terms of the cumulants is given by 
\be
P(\delta)d\delta=\frac{d\delta}{2\pi i}\frac{1}{\sigma_{\delta}^2}\int_{-i\infty}^{i\infty}dy\exp\left[\frac{y\delta}{\sigma_{\delta}^2}-\frac{\mathcal{S}(y)}{\sigma_{\delta}^2}\right]
\label{cumulantPDF}
\ee
where $\sigma_{\delta}^2$ is the variance and the generating function $\mathcal{S}(y)$ is defined by
\be
\mathcal{S}(y)=\sum_{n=2}^{\infty}\mathcal{S}_n\frac{(-1)^{n-1}}{n!}y^n\;.
\label{Ssum}
\ee
Changing integration variables to $\hat{y}=y/\sigma_{\delta}$, this is
\be
P(\delta)d\delta=\frac{d\delta}{\sigma_{\delta}}\frac{1}{2\pi i}\int_{-i\infty}^{i\infty}d\hat{y}\exp\left[\hat{y}\left(\frac{\delta}{\sigma_{\delta}}\right)-\sum_{n=2}^{\infty}\frac{(-1)^{n-1}}{n!}\hat{y}^n(\mathcal{S}_n\sigma_{\delta}^{n-2})\right]
\label{cumulantPDF2}
\ee
Using the saddle point approximation to perform the integral in Eq.(\ref{cumulantPDF}) and collecting terms of the same order, one arrives at the Edgeworth expansion:
\be
P(\nu)d\nu=\frac{d\nu}{\sqrt{2\pi}}e^{-\nu^2/2}\left[1+\sigma_{\delta}\frac{\mathcal{S}_3}{6}H_3(\nu)+\sigma_{\delta}^2\left(\frac{\mathcal{S}_4}{24}H_4(\nu)+\frac{\mathcal{S}_3^2}{72}H_6(\nu)\right)+\dots\right]
\label{Edgeworth}
\ee
where $\nu=\delta/\sigma_{\delta}$ and the $H_n$ are Hermite polynomials. Both Eq.(\ref{cumulantPDF2}) and Eq.(\ref{Edgeworth}) are an expansion in what we will call the dimensionless cumulants, $\mathcal{S}_n\sigma^{n-2}$, so truncations of these expressions give a good approximation of the actual PDF if 
\be
1\gg |\mathcal{S}_3|\sigma\gg|\mathcal{S}_4|\sigma^2\dots
\ee
where we have drop the subscript to indicate that this expression is general. If only the first few terms in the expansion are kept the recovered probability distribution will be approximate and have a limited range of validity. For example, the PDF can develop negative regions if only the first $\mathcal{S}_3$ term is kept.  This discussion has been one-dimensional, but we are often concerned with spherically symmetric situations where the fluctuations are smoothed on a scale $R$. Then all of the quantities above become functions of $R$, but otherwise the expressions are unchanged. Of course, Eq.(\ref{Edgeworth}) is also an expansion in $\nu$, so the regime of validity of the Edgeworth expansion for large scale structure calculations also depends on the smoothing scale (which affects $\sigma_{\delta}$) and the region of the PDF of interest ($\delta$).

An observable like the number of large structures (galaxy clusters) depends on the probability of a several-sigma fluctuation, and so is sensitive to the area under the tail of the PDF. The potential to constrain a fundamental model of non-Gaussianity by counting structures relies on the validity of expanding in the dimensionless cumulants, or on some other more exact knowledge of the relevant distribution. 

\subsection{Hierarchical structure and the local model}
The local model ansatz for primordial non-Gaussianity is \cite{Salopek:1990jq}
\be
\zeta({\bf x})=\zeta_G({\bf x})+\frac{3}{5}f_{NL}\left[\zeta_G^2({\bf x})-\langle\zeta_G^2({\bf x}) \rangle\right]
\label{eq:NG1}
\ee
where $\zeta({\bf x})$ is the primordial curvature perturbation, $\zeta_G({\bf x})$ is a Gaussian random field ($\langle\zeta_G^2({\bf x}) \rangle\equiv\sigma^2$) and the amplitude of non-Gaussianity is parameterized by $f_{NL}$, which we take to be constant for simplicity. This generates correlation functions with hierarchical scaling. For the two-point function at a single point in real space to be nearly given by the two-point of the Gaussian piece requires
\be
2\left(\frac{3}{5}f_{NL}\right)^2\sigma^4\ll\sigma^2\Rightarrow\frac{3}{5}|f_{NL}|\sigma=\frac{1}{6}|\mathcal{S}_3|\sigma\ll 1\;.
\ee
This is the condition that the model is not `too' non-Gaussian. (Numerically, on CMB scales, $\langle\zeta_G^2({\bf x}) \rangle\equiv\sigma^2\approx10^9$, so this gives $|f_{NL}|\ll10^{9/2}$, which is much weaker than the current observational constraint.) Furthermore, higher order cumulants go like powers of $f_{NL}\sigma$, so $\mathcal{S}_n\propto (f_{NL}\sigma)^{n-2}$. As long as $|f_{NL}|\sigma\ll1$, the expansion in cumulants is valid. 

Non-Gaussianity of the local shape is a feature of multi-field models, but in such scenarios Eq.(\ref{eq:NG1}) is only the first term in a more general expansion
\be
\zeta({\bf x})=\zeta_G({\bf x})+\frac{3}{5}f_{NL}\left[\zeta_G^2({\bf x})-\langle\zeta_G^2({\bf x}) \rangle\right] +g_{NL}\zeta_G^3+\dots
\ee
The range of possibilities for non-Gaussianity in multi-field models is just beginning to be studied systematically, but as suggested in \cite{Enqvist:2008gk} for a curvaton scenario, such models are not necessarily hierarchical. Even if each individual field is perturbative and generates hierarchical correlations, a fine tuning may allow the skewness of the observed curvature perturbation to be nearly zero while the kurtosis is measurably large. In addition, although $\mathcal{S}_3$ is constrained by observation and $\mathcal{S}_4$ is in principle, it has not yet been observationally established if $|\mathcal{S}_3|\sigma\gg|\mathcal{S}_4|\sigma^2$. 

\subsection{Fluctuations and effective field theory}
This section has so far taken a largely observational point of view, which we now want to connect to the scalar field theory during inflation. For inflation from a single scalar field, non-Gaussianity larger than that generated by the usual slow-roll must come from non-trivial kinetic terms (as long as the potential is smooth and nearly flat). Using the parameter $f_{NL}$ as a proxy for magnitude of non-Gaussianity, with $|f_{NL}|>5$ probably observable in the near future, one might guess that $|f_{NL}|\sim1$ for general theories of the type Eq.(\ref{fluidact}). The reasoning is the following \cite{Creminelli:2003iq}: Consider a Lagrangian $P(X,\phi)$ with $\phi$ the inflaton and $X=-\frac{1}{2}g^{\mu\nu}\partial_{\mu}\phi\partial_{\nu}\phi$. Suppose
\be
P(X,\phi)=-V(\phi)+X+a_1\frac{X^2}{M^4}+a_2\frac{X^3}{M^8}+\dots
\ee
where the potential energy $V(\phi)$ will drive inflation and $M$ is some scale (like $M_p$ or the string scale $m_s$) suppressing higher order operators. If $X/M^4\ll1$, this series can be truncated. Then if $X/M^4\ll1$ and the coefficients $a_i$ are of order 1, $P_{,X}\sim 1$, $2XP_{,XX}\ll1$ and $c_s^2\sim\mathcal{O}($a few$\times10^{-1})$. The sound speed was defined in Eq.(\ref{definecs}) and gives the magnitude of the non-Gaussianity through $|f_{NL}|\propto c_s^{-2}$ \cite{Chen:2006nt}. This is easy to see by examining the ratio of terms in the Lagrangian that are cubic in fluctuations to the free, quadratic part.

Here we will be interested in the case $c_s^2\ll1$. To see how this regime can be under control, we should relate the scale $M$ to other parameters in the theory, starting with the relation
\be
\dot{\phi}=-\frac{2M_p^2H^{\prime}}{P_{,X}}
\ee
where primes are derivatives with respect to $\phi$ and over-dots are time derivatives, and rewriting in terms of the parameter $\epsilon$
\be
\epsilon=-\frac{\dot{H}}{H^2}=\frac{2M_p^2}{P_{,X}}\left(\frac{H^{\prime}}{H}\right)^2
\ee
we find
\be
\frac{X}{M^4}=\frac{\epsilon H^2M_p^2}{P_{,X}M^4}
\label{howbig}
\ee
With $P_{,X}\approx 1$, this gives
\be
\frac{X}{M^4}\ll1\Rightarrow \epsilon H^2M_p^2\ll M^4
\ee
As pointed out in \cite{Weinberg:2008hq}, for $M\sim \sqrt{\epsilon}M_p$, the COBE normalization implies $\frac{H}{M}\lesssim4\times10^{-4}$. In addition, the condition that the kinetic energy is small compared to the potential energy is $X\ll 3M_p^2H^2\Rightarrow\frac{X}{M^4}\ll\frac{3}{\epsilon}\left(\frac{H}{M}\right)^2$, which is small unless $\epsilon$ is unusually tiny. 

In contrast, we are interested in the case where $\frac{X}{M^4}$ is not too small so that the derivative interactions are significant. This is more natural when both the Hubble scale and the scale $M$ are well below the Planck scale (see Eq.(\ref{howbig})). Notice that for inflation to continue (that is, for potential energy to remain dominant) in the regime where $\frac{X}{M^4}$ is close to one, we must have $V(\phi)>X\sim M^4$. (Although we require that the inflaton mass, $m$, is less than $M$.)

With these relations in mind, we now turn to a consideration of correlation functions in actions of the fluid type, Eq.(\ref{fluidact}).

\subsection{Structure of tree-level correlation functions for a perfect fluid}
To examine the structure of correlation functions for actions of the $P(X,\phi)$ type, consider Taylor expanding the action in Eq.(\ref{fluidact}) around the background value of the field $\phi_0$, ordering terms according to powers of the fluctuation so that
\be
S=S_0+S_2+S_3+\dots
\ee
where $S_0$ is the background solution, $S_1$ vanishes when the equations of motion are satisfied, $S_2$ is quadratic in fluctuations and is used to write the mode equations for the fluctuations. The solution to those equations is used to evaluate the 3-point function coming from terms in $S_3$ and to give a magnitude of non-Gaussianity. We can then determine the interaction Hamiltonian $H_I$, starting with the terms from $S_3$, and start computing correlation functions as outlined in \cite{Weinberg:2005vy}. Correlation functions at time $t$, taken to be long after horizon exit but before the end of inflation, are computed using the ``in-in" formalism \cite{Calzetta:1986cq, Maldacena:2002vr, Weinberg:2005vy}. At tree level, 
\ba
&&\Expect{\delta\phi(k_1,t)\dots\delta\phi(k_n,t)}=\\\nonumber
&&i\int_{-\infty}^{t}\;dt^{\prime}\;\Expect{[H_I^{(n)}(t^{\prime}),\delta\phi(k_1,t)\dots\delta\phi(k_n,t)]}
\label{keldysh}
\ea
where $H_I^{(n)}$ is the term with $n$ powers of the fluctuation. As reviewed in Appendix \ref{Honest}, performing this calculation for an action including an $X^2$ term shows that the dominant contribution to the three-point correlation of the primordial curvature (when $c_s$ is small) has the form
\ba
\Expect{\zeta(\vec{k}_1)\zeta(\vec{k}_2)\zeta(\vec{k}_3)}&=&(2\pi)^3\frac{\delta^3(\vec{k}_1+\vec{k}_2+\vec{k}_3)}{(k_1k_2k_3)K^3}\\\nonumber
&&\times \left(\frac{\mathcal{P}_{\zeta}^2}{c_s^2}\right)B(\vec{k}_1,\vec{k}_2,\vec{k}_3)
\ea
where $B(\vec{k}_1,\vec{k}_2,\vec{k}_3)$ is dimensionless with terms of order 1 and terms suppressed by $c_s^2$, and $K$ is the sum of the momenta.

Since we are just interested in how higher order correlation functions scale with the amplitude $\mathcal{P}_{\zeta}$ and the sound speed, we do not actually need to do the full calculation for more complicated actions. Instead, we can Taylor expand the action $P(X,\phi)$ \cite{Huang:2006eh, Leblond:2008gg}. The Lagrangian density is used below to simplify notation,
\ba
a^{-3}\mathcal{L}_2&=&\frac{1}{2}P_{,X}[\dot{\df}^2-a^{-2}(\nabla\df)^2]+\frac{1}{2}\dot{\phi}^2\dot{\df}^2P_{,XX} + \dots\\\nonumber
a^{-3}\mathcal{L}_3&=&\frac{1}{2}P_{,XX}\dot{\phi}\dot{\df}[\dot{\df}^2-a^{-2}(\nabla\df)^2] +\frac{1}{6}P_{,XXX}\dot{\phi}^3\dot{\df}^3 +\dots\\\nonumber
a^{-3}\mathcal{L}_4&=&\frac{1}{8}P_{,XX}[\dot{\df}^2-a^{-2}(\nabla\df)^2]^2+\frac{1}{4}P_{,XXX}\dot{\phi}^2\dot{\df}^2[\dot{\df}^2-a^{-2}(\nabla\df)^2]\\\nonumber
&&+\frac{1}{24}P_{,XXXX}\dot{\phi}^4\dot{\df}^4+\dots
\ea
Two simplifying relations we use here and throughout the paper are
\ba
\label{roughderivs}
\frac{\partial}{\partial t} \sim H\; , & & \frac{\partial}{\partial x} \sim \frac{a H}{c_s}\; .
\ea
Then, using the rms value of fluctuations at horizon crossing, inserting powers of $\dot{\phi}^2=2X$ as needed and identifying the combination $\mathcal{P}_{\zeta}=\frac{H^2}{M_p^2\epsilon c_s}$ gives:
\ba\label{higher}
\mathcal{L}_2&=&a^3\frac{H^4}{c_sP_{,X}}\left[P_{,X}\left(1-\frac{3}{c_s^2}\right)+2XP_{,XX}\right]+ \dots\\\nonumber
&=&-2a^3\Sigma \mathcal{P}_{\zeta}+\dots\\\nonumber
\mathcal{L}_3&=&a^3\frac{1}{2}\left(\frac{2H^4}{\dot{\phi}^2c_sP_{,X}}\right)^{3/2}\left[X^2P_{,XX}\left(1-\frac{3}{c_s^2}\right)+\frac{2}{3}X^3P_{,XXX}\right] +\dots\\\nonumber
&&=-\mathcal{L}_2\frac{P^{1/2}_{\zeta}}{2c_s^2}\left[3(1-c_s^2)-\frac{2\lambda c_s^2}{\Sigma}\right]+\dots\\\nonumber
\mathcal{L}_4&=&\frac{a^3}{2}P^2_{\zeta}\left[X^2P_{,XX}\left(1-\frac{3}{c_s^2}\right)^2+4X^3P_{,XXX}\left(1-\frac{3}{c_s^2}\right)+\frac{4}{3}X^4P_{,XXXX}\right]+\dots\\\nonumber
&=&-\mathcal{L}_2\frac{9P_{\zeta}}{8c_s^4}+\dots
\ea
where two useful combinations of derivatives have been defined
\ba
\label{SigLam}
\Sigma&=&XP_{,X}+2X^2P_{,XX}=\frac{XP_{,X}}{c_s^2}=\frac{\epsilon H^2M_p^2}{c_s^2}\\\nonumber
\lambda&=&X^2P_{,XX}+\frac{2}{3}X^3P_{,XXX}\; .
\ea
As long as the coefficients of higher order terms are not too large (conditions like $\lambda\lesssim3\Sigma/2c_s^2$), this pattern continues and the dominant contributions at order $n$ in the small sound speed limit go like
\be
H_I^{(n)}\sim A_n\int d^3x\;a^3\mathcal{L}_2\left(\frac{\mathcal{P}_{\zeta}^{1/2}}{c_s^2}\right)^{n-2}
\label{HI}
\ee
where $A_n$ is some constant coefficient. Exceptions to this behavior require fine-tuning, and an explicit example was worked out in \cite{Engel:2008fu}.

We can now estimate the integral in Eq.(\ref{keldysh}) using $(\Delta x)^3\;\Delta t\sim(c_s/aH)^3H^{-1}$ to see that $\Delta t(\Delta x)^3a^3\mathcal{L}_2\sim1$. This estimate assumes that the dominant physical contribution to the integral should be near the scale of horizon crossing. Then the $n$-point correlation scales like 
\be
\Expect{\zeta^n}\propto\Expect{\zeta^2}^{n-1}(c_s^{-2})^{n-2}
\ee
This is the hierarchical form, but with extra powers of the sound speed. To check if those factors are problematic, we note that the dimensionless combination of the magnitude of the $n$-point function scaled by the two-point function, the dimensionless cumulant, goes like:
\be
\mathcal{S}_n\sigma^{n-2}=\frac{\Expect{\zeta^n}}{\Expect{\zeta^2}^{n-1}}\Expect{\zeta^2}^{(n-2)/2}\propto A_n\left(\frac{P_{\zeta}^{1/2}}{c_s^2}\right)^{n-2}
\ee
So for $c_s^4\gg P_{\zeta}$ and for $A_n$ of order one, the truncated cumulant expansion will give a reasonable approximation to the PDF. 

\section{Expanding the most general action with higher derivatives}
\label{expanding}
The observationally interesting case of large non-Gaussianity (small sound speed) requires that $X/M^4$ be close enough to one that many terms in the expansion are important. Although there are special examples like the DBI action where the entire structure of first-derivative terms is known, it is interesting to ask how small a sound speed is reasonable in the most general single field inflation. The procedure above indicates the perturbative approach is valid while $c_s^4>P_{\zeta}$, but considered only single derivative terms. In \cite{Cheung:2007st}, the same conclusion was reached considering first derivative terms organized in different way. That work, an effective theory of fluctuations, considered higher derivative terms in operators depending on the extrinsic curvature and its derivatives. The approach of \cite{Cheung:2007st} is very useful for understanding a variety of cases, but in most inflationary models it is the effective theory of the inflaton itself that is uncovered term by term. With that in mind, we consider the first few steps of an expansion of the action in powers of derivatives.

The generic expansion to fourth order in derivatives was recently carried out by Weinberg \cite{Weinberg:2008hq}, who found that the only new term generically present in the action\footnote{At four derivatives, this is the only important term for scalar fluctuations even after terms involving metric derivatives are considered. We are not repeating the gravitational part of the analysis for six derivative terms, but we do expect some terms involving the metric to remain, suppressed by $M/M_p$.} has the form $f_{1}(\phi)\frac{X^2}{M^4}$. To arrive at this simple answer, the quadratic equation of motion was used to replace $\Box\phi=V^{\prime}(\phi)$, which ensures that there are no extra propagating degrees of freedom.

\subsection{Expansion at six derivatives}
\label{sixderiv}
From an effective field theory point of view, one non-renormalizable operator cannot be added to the theory in isolation. Continuing to six-derivative terms and using integration by parts to remove redundancies, there are eight independent terms involving derivatives of the scalar field (ignoring new terms involving derivatives of the metric, which are likely to be suppressed by powers of $M_p$ rather than $M$). They have the following structure:
\ba
&&X^3\;,\;X^2\Box\phi\;,\;X(\Box\phi)^2\;,\;(\Box\phi)^3\;,\;\phi\Box^3\phi\;,\;Xg^{\mu\nu}(\partial_{\mu}\phi)(\partial_{\nu}\Box\phi)\;,\\\nonumber
&& g^{\mu\nu}g^{\alpha\beta}(\partial_{\mu}\phi)(\partial_{\alpha}\phi)(\partial_{\nu}\partial_{\beta}\Box\phi)\;,\;g^{\mu\nu}g^{\alpha\beta}(\partial_{\mu}\partial_{\alpha}\phi)(\partial_{\nu}\partial_{\beta}\phi) X\;.
\ea
Notice that the only term involving two fields may be written as powers of the box operator, and that there are only two terms that do not involve box.

In the expansion up to four-derivative terms, using the lowest order equation of motion in the Lagrangian conveniently removed second derivatives of $\phi$ and recovered the simple structure of the $\phi$ propagator \cite{Weinberg:2008hq}. Since at any order in derivatives, the only term that is quadratic in $\phi$ can be written as $\phi\Box^n\phi$, eliminating box in this simple way ensures that the $\phi$ propagator keeps a single physical pole, with corrections in powers of $m/M$. Substituting for the box operator in the same way in the six derivative terms adjusts the coefficients of lower order terms in the action. Although we will generate higher order derivative terms at each stage that will {\it not} be removed, these are still small perturbations and so we do not worry about having additional degrees of freedom. Then, up to terms with six derivatives, we may write
\ba
\label{uptosix}
\mathcal{L}&=&\sqrt{-g}\left[\frac{M_p^2}{2}R+X-V(\phi)+4f_1(\phi/M)\frac{X^2}{M^4}+2f_2(\phi/M)\frac{X}{M^6}g^{\alpha\beta}g^{\mu\nu}\partial_{\alpha}\partial_{\mu}\phi\partial_{\beta}\partial_{\nu}\phi\right.\\\nonumber
&&\left.+8f_3(\phi/M)\frac{X^3}{M^8}+F(R,\phi)\right]
\ea
where the $f_i(\phi/M)$ are dimensionless, order one functions and the last term is a placeholder for terms involving more than two derivatives, with at least some derivatives acting on the metric and scaling with $M_p\gg M$. We will examine the contribution of the two new six-derivative terms to the correlation functions in the next section. Note that the sound speed defined in Eq.(\ref{definecs}) is defined from the background solution and so does not receive a significant correction from the higher derivative terms.

Before we continue, we comment that one can imagine being more rigorous about eliminating higher derivative contributions. For a classical point particle Lagrangian, there is a systematic procedure for obtaining a Hamiltonian $H(\phi)$ by replacing higher derivatives by functions of the field and its first derivative as long as the higher derivatives can be considered perturbations \cite{Jaen:1986iz, Simon:1990ic}. This ensures that the only solutions to the equations of motion are consistent with perturbative corrections to the free field solutions, and only two initial conditions are needed. This idea is built in to the field theory structure that requires the interaction part of the Hamiltonian to be in some sense small, and in the next section we will find the answer suggested by the point particle example - as long as higher derivative terms enter as perturbations there is a regime where their contribution, both to the background and to correlation functions of fluctuations, is small compared to single-derivative terms, even if one has reason to keep an entire power series of single derivative terms.

\section{Estimating the higher derivative contributions}
\label{pert}
Truncating the action $P(X,\phi)$ assumes that $X/M^4\ll 1$. Any truncation obviously leads to a truncation in correlation functions. Including terms up to $(X/M^4)^n$ (and ignoring terms from the potential, which are less important) generates correlations up to $\Expect{(\delta\phi)^{2n}}$. We saw above that $P_{\zeta}<c_s^4$ was required for the cumulants to be ordered in the case that only first derivatives are considered. Expanding a more general action up to $p$ derivatives allows one to calculate up to $\Expect{(\delta\phi)^p}$. We can estimate the relative size of the various contributions to, say the three point function. Replacing $\phi=\phi_0+\delta\phi$ in the action Eq.(\ref{uptosix}) and collecting terms with three powers of the fluctuation, we have
\ba
S_3(\phi)&=&\int d^4x \sqrt{-g}\left\{4f_1(\phi/M)\frac{\dot{\phi_0}}{M^4}[(\dot{\delta\phi})^3-a^{-2}(\dot{\delta\phi})(\nabla\delta\phi)^2]\right.\\\nonumber
&&+2f_2(\phi/M)\frac{1}{M^4}\left[\frac{\dot{\phi_0}}{M^2}(\dot{\delta\phi})[(\ddot{\delta\phi})^2-2a^{-2}(\nabla\dot{\delta\phi})^2+a^{-4}(\partial_i\partial_j\delta\phi)(\partial^i\partial^j\delta\phi)]\right.\\\nonumber
&&\left.+\frac{\ddot{\phi_0}}{M^2}(\ddot{\delta\phi_0})[(\dot{\delta\phi})^2-a^{-2}(\nabla\delta\phi)^2]\right]\\\nonumber
&&\left.+8f_3(10\phi/M)\frac{\dot{\phi_0}}{M^4}\frac{X}{M^4}[5(\dot{\delta\phi})^3-3a^{-2}(\dot{\delta\phi})(\nabla\delta\phi)^2]+\dots\right\}
\ea
Using $\partial/\partial t\sim H$ and $\partial/\partial x\sim aH/c_s$, we immediately see that the contributions to the three point function with the strongest $c_s$ dependence go like
\be
H_I^{(3)}\propto\int dt\;a^3(\delta\phi)^3\frac{\dot{\phi_0}}{M^4}\frac{H^3}{c_s^2}\left[4f_1+2f_2\frac{H^2}{M^2c_s^2}+24f_3\frac{X}{M^4}\right]
\label{thirdorder}
\ee
In the limit of very small sound speed, $X/M^4\rightarrow1$ and so the contributions from all powers of $X/M^4$ in the action will be equally important for the correlation functions. These contributions can be collected by writing the action as a Taylor expansion, as in \cite{Leblond:2008gg} or implicitly in \cite{Cheung:2007st}. We also see that while $H/M<c_s$, contributions from higher derivative terms are suppressed relative to single derivative terms. This assuming no parametric dependence on $H/M$, etc in the coefficient, which in any case is likely to further suppress the term.

When is $H/M<c_s$? We have assumed that $X/M^4<1$. If the previously derived condition $P_{\zeta}<c_s^4$ also holds, we see that
\ba
\frac{X}{M^4}&=&\frac{\epsilon H^2M_p^2}{P_{,X}M^4}<1\\\nonumber
\Rightarrow\frac{H^4}{M^4}&<&(P_{,X}c_s)\frac{H^2}{M_p^2\epsilon c_s}<(P_{,X}c_s)c_s^4
\ea
For DBI, $P_{,X}c_s=1$, and we have exactly $H/M<c_s$. The expansion to six derivative terms as written above gives $P_{,X}c_s<1$, and it has been argued that in more general cases one can set $P_{,X}c_s=1$ as a gauge choice (observations are not sensitive to $P_{,X}$ alone.) \cite{Bean:2008ga}.  In the perturbative case then, correlation functions of a general scalar field action are dominated by contributions from the $P(X,\phi)$ part, assuming the coefficients of powers of $X/M^4$ are the renormalized coefficients, and those correlations are hierarchical. When the perturbative condition breaks down, not only are the dimensionless cumulants from single derivative terms no longer a converging series, but all higher derivative corrections are equally important. Continuing to more derivatives in the action can only bring in more copies of $X/M^4$ or $H/Mc_s$, so, up to gravity terms, we are done. We see that $X/M^4<1$ and $P_{\zeta}<c_s^4$ imply $H/M<c_s$, and as both the first two inequalities are both saturated, so is the third.

\section{Example: the DBI action}
\label{DBI}
\subsection{The lowest order brane action in an AdS background}
There is a very interesting scenario with $c_s\ll 1$ that is a useful test case for effective field theory ideas (and where one might hope to study the non-pertrubative regime of very small $c_s$). The Dirac-Born-Infeld (DBI) action contains an infinite series of powers of $X/M^4$ summed to a square-root that enforces $X/M^4\leq1$. In the case of a $D$3 brane moving down a warped throat \cite{Silverstein:2003hf, Alishahiha:2004eh}, $M$ is the warped string scale, where the warp factor depends on the position of the brane in the background AdS geometry. The position-dependent warp factor translates into a scale $M$ that decreases during inflation and a scale-dependent non-Gaussian signature in the resulting density perturbations (which may be observable \cite{LoVerde:2007ri}). The radial brane position is $r$, with $r_{min}<r<R$ where $R$ (large in string units) is the scale of the throat and $r_{min}$ is the parameter giving the minimum scale where the throat smoothly ends. Then (a very simplified version of) the lowest order brane action is
\be
S=-\int d^4x\; a^3\left[T_3h(r)\sqrt{1-\dot{r}^2h^{-1}(r)} -T_3h(r)+V(\phi)\right]
\ee
Here $T_3=[(2\pi)^3g_s\alpha^{\prime 2}]^{-1}$ is the three-brane tension in terms of the string coupling $g_s<1$ and the string length $\alpha^{\prime}\propto l_s^2=m_s^{-2}$. We have assumed the 10-dimensional metric takes the form\footnote{This has not been shown to follow consistently as a supergravity solution, and there are some indications that something is missing in this description \cite{Becker:2007ui, Chen:2008hz}. However, that is not likely to be relevant for the point we are making here.}. 
\be 
\label{10Dwarp} ds_{10}^2 =
h^{-1/2}(y) g_{\mu \nu}\,dx^{\mu} dx^{\nu}+ h^{1/2}(y)
g_{mn}dy^mdy^n\; ,
\ee
where the warp factor $h$ depends only on the coordinates $y$ of the extra dimensions and the four-dimensional metric $g_{\mu\nu}$ is Friedmann-Robertson-Walker with scale factor $a$.

We consider only potential energy dominated cases, so that the Hubble parameter during inflation $H=\dot{a}/a$ is given by $3M_p^2H^2\approx V(\phi)$ to a good approximation. For a purely AdS background, the warp factor is $h(r)=R^4/r^4$, although the throat may smooth out and have a region of constant $h(r)=h_0$ near the minimum $r_{min}$ \cite{Klebanov:2000hb}. The canonical inflaton is $\phi=\sqrt{T_3}r$, so that the action in terms of $\phi$ (and restoring covariance) is
\be
S=-\int d^4x\; a^3\left[f(\phi)\sqrt{1-2Xf^{-1}(\phi)}-f(\phi)+V(\phi)\right]
\label{DBIlowest}
\ee
where $f(\phi)\equiv T_3h^{-1}(\phi)=M^4(\phi)$. 
In this case, the $M$ decreases as $\phi$ does, and the sound speed decreases like $c_s\propto\phi^2/M_p^2$ \cite{Silverstein:2003hf}. This is plotted below.
\begin{figure}[h]
\begin{center}
\includegraphics[width=0.5\textwidth,angle=0]{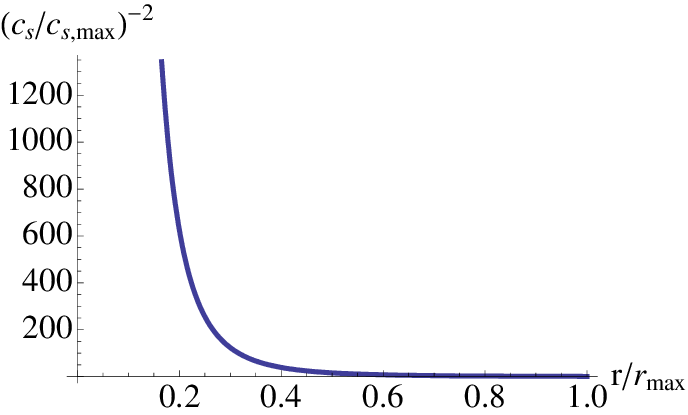} 
\caption{For a brane moving in an AdS background geometry, the sound speed continually decreases until the geometry changes near $r_{min}$. If $c_s$ gets too small ($c^{-2}_s>\mathcal{P}^{-1/2}_{\zeta}\approx10^{9/2}$), the perturbative calculation of correlation functions based on the lowest order brane action is no longer valid.}
\label{csgrows}
\end{center}
\end{figure}
So, at this level, if the ratios $H/M_p$ (which sets the magnitude of $\mathcal{P}_{\zeta}$) and $r_{min}/R$ (which determines the minimum $c_s$) are taken to be independent (i.e. controlled by separate microphysical parameters) it looks like one may run into a regime with $c_s^4<\mathcal{P}_{\zeta}$, in which case the effective theory based on the lowest order action breaks down \cite{Leblond:2008gg}. Of course, we might hope that the stringy picture indicates what physics is missing, and it may be that inflation continues even past this point. We will use the discussion of higher derivative terms to pursue this idea a little further.

\subsection{Higher order corrections}
Eq.(\ref{DBIlowest}) is the lowest order action for the brane. Corrections to the fluctuation spectrum that depend on higher derivatives of the position can be organized as the extrinsic curvature, $K_{\mu\nu}$, and its derivatives. In fact, this is a useful organization in the general theory of fluctuations \cite{Cheung:2007st}. The extrinsic curvature of the brane due to fluctuations on the homogeneous background ($\partial_i\phi_0=0$) is
\be
K_{\mu}^{\mu}\approx\frac{\partial_i^2(\delta\phi)}{a^2M^2}\;
\ee
where we have dropped small corrections due to the variation in the warp factor with $\phi$. We can verify that in the perturbative regime the curvature is smaller than the local warped string scale since
\be
K_{\mu}^{\mu}\approx\left(\frac{H}{Mc_s}\right)^2\frac{H}{2\pi}<M.
\label{extrinsic}
\ee
as long as $H/M<c_s$, $c_s<1$. This is good,  since if the radius of curvature was less than $M^{-1}$) the brane would be unstable to decay by self-annihilation in the fluctuating regions. Eq.(\ref{extrinsic}) also demonstrates that the gradient energy is not yet so large that inflation ends, since $M^4<V(\phi)$. However, it is an interesting dynamical question whether or not caustics may develop on a time scale much shorter than the Hubble time \cite{Saremi:2004yd}. To address that question accurately, we would need to know the form of higher derivative terms in the DBI action (although we note that self-annihilation of the brane due to folding cannot be whole answer, if it happens at all, since $D$3 charge must be globally conserved).

We can speculate that the DBI action in a background that gives rise to inflation is corrected by higher derivative terms as
\be
S_{DBI, \text{guess}}=-\int d^4x\; a^3\;f(\phi)\sqrt{1-2X/M^4}\times\left[1+\mathcal{F}^{mnkl}(X/M^4)(\partial_m\partial_n\phi)(\partial_k\partial_l\phi)+\dots \right]\;.
\label{DBIcorrect}
\ee
If the first term in $\mathcal{F}(X/M^4)$ is proportional to $X/M^4$ and contains the index contraction we used in Eq.(\ref{uptosix}), this guess resembles what would be obtained by T-duality from the known corrections in derivatives of the field strength in the bosonic action (the first correction to the superstring action is at eight derivatives) \cite{Andreev:1988cb}. The multiplication by the square root forces the entire derivative structure to vanish as $c_s\rightarrow0$. In addition, if all higher derivative terms enter multiplied by powers of $X$ (as suggested by the form of field strength derivative terms), the higher derivative structure vanishes when the velocity is zero. 

The action suggested by brane dynamics naturally has a more restricted structure than the EFT structure considered in \cite{Cheung:2007st}. There the most general effective action for the fluctuations of a single scalar field is organized into contributions to the $n$-point function from a summation of operators with powers of $X/M^4$, and, separately, the extrinsic curvature (higher derivative) terms. The extrinsic curvature terms were assumed to be independent of the terms proportional to powers of $X$. Here the symmetries that control the square-root summation (conformal invariance, supersymmetry) also control the structure of higher derivative terms. More generally, if we believe we know how to sum all (or enough of) the operators contributing to, say, the three point to understand a measurably large non-Gaussianity, we may also know the structure relating those terms to the extrinsic curvature terms.

The action in Eq.(\ref{DBIcorrect}) is still not complete, and in particular does not contain terms with curvature corrections or couplings to the background fields generating the brane potential. To answer the question of what happens for brane positions so far in the IR end that $c^4_s<P_{\zeta}$, we must consistently include these other corrections. Since this is specific to the string theory scenario, we will not pursue it further here.

\section{Conclusions}
\label{conclude}
Future measurements of non-Gaussianity (even with a null result) can provide unique information about physics at the highest energy scale accessible, but the utility of that information depends on our ability to translate fundamental models to observation. A feature of that translation, beyond the shape and amplitude of the three-point function, is the expected structure of correlation functions. Here we have demonstrated explicitly a very natural result: that the most general, perturbative, single-field inflation models that generate large non-Gaussianity from higher derivative terms have a hierarchical structure. That is, as long as the sound speed is large enough that perturbative calculations can be trusted, the statistics can be captured by expanding in moments around a Gaussian distribution. The perturbative condition (which assumes $X/M^4\lesssim1$) takes two useful forms:
\be
c_s^4>P_{\zeta}\;\;\text{-or-}\;\;\frac{H}{M}<c_s
\ee
Derivative interactions are relevant (and non-Gaussianity significant) if the scale of the effective theory is not too far above the Hubble scale. While it may seem fine-tuned to consider this coincidence of scales (rather than $M\sim M_p$), it is sometimes a trade-off from fine-tuning the flatness of the inflaton potential (which was the context in which this effect was originally introduced \cite{Silverstein:2003hf}). These results assume order one coefficients for all terms in the action, so fine tuned single field or multi-field examples can be exceptions. A multi-field model that is not hierarchical is given in \cite{Enqvist:2008gk}, and the fine-tuning for single field examples is discussed in \cite{Engel:2008fu}.

A significantly large non-Gaussianity is easiest to explain if we know something about the UV structure of the theory. One of the few known examples is based on the DBI action for a $D$-brane, where higher derivative corrections may also be calculated. In that case, it is likely that the same symmetries that lead to the square-root summation of first derivative terms also control the at least some of the structure of higher derivatives.

Investigating the behavior of a general scalar field shows that actions of the $k$-inflation type (actions of the $P(X,\phi)$ type) are not sufficiently general beyond four derivative terms, although the difference in the perturbative regime is very small for the background solution and down by a factors of $H/Mc_s$ in the correlation functions. Interestingly, a simple extension of $P(X,\phi)$ actions to multiple fields is also modified at four derivatives and in that case the deviation appears to be more observationally important \cite{Langlois:2008qf}.

\acknowledgments
This work has benefitted significantly from conversations with Puneet Batra, Louis Leblond, Eugene Lim, Marilena LoVerde, Alberto Nicolis, and David Seery. This work was supported by the DOE under DE-FG02-92ER40699.

\appendix
\section{Calculating correlation functions}
\label{Honest}
In this section we calculate the $\delta\phi$ correlations and use the first order gauge transformation to find the leading contribution to correlations of $\zeta$ (the quantity that is constant outside the horizon). This is by no means a complete calculation of $\zeta$ correlations, but it does correctly capture the behavior of the leading term in the small $c_s$ limit. This can be checked against the calculation of \cite{Arroja:2008ga}, which works out the complete trispectrum for a general sound speed model. 

\subsection{Statistics of the action expanded to four-derivative terms}
Weinberg \cite{Weinberg:2008hq} has argued that the general form of the effective action up to terms with four derivatives, and for the moment ignoring gravity, is
\ba
\mathcal{L}&=&\sqrt{-g}\left[X-V(\phi)+4f_1(\phi)\frac{X^2}{M^4}+\dots\right]
\ea
where $f_1(\phi)$ is a dimensionless function of $\phi$ and $X=-\frac{1}{2}g^{\mu\nu}\partial_{\mu}\phi\partial_{\nu}\phi$. Expanding in terms of the fluctuation $\pi=\delta\phi/\dot{\phi}_0$ and labeling the background piece $\mathcal{L}_0$, this is
\ba
\mathcal{L}&=&\mathcal{L}_0+a^3M_p^2\dot{H}(-\dot{\pi}^2+a^{-2}(\vec{\nabla}\pi)^2)\\\nonumber
&&+\frac{16a^3M_p^4\dot{H}^2f_1(\phi)}{M^4}\left(\dot{\pi}^2+\dot{\pi}^3-\frac{\dot{\pi}(\vec{\nabla}\pi)^2}{a^2}+\frac{\dot{\pi}^4}{4}-\frac{\dot{\pi}^2(\vec{\nabla}\pi)^2}{2a^2}+\frac{(\vec{\nabla}\pi)^4}{4a^4}\right)+\dots
\ea
From this, we can compute the tree level three and four point functions of $\pi$. Quantizing the fluctuations
\be
\delta\phi(\vec{x},\eta)=\int\frac{d^3k}{(2\pi)^3}[u(\vec{k},\eta)a(\vec{k})e^{i\vec{k}\cdot\vec{x}}+u^{*}(\vec{k},\eta)a^{\dagger}(\vec{k})e^{-i\vec{k}\cdot\vec{x}}]
\ee
where $[a(\vec{k}),a^{\dagger}(\vec{k}^{\prime})]=(2\pi)^3\delta^3(\vec{k}-\vec{k^{\prime}})$ and from the quadratic terms
\be
u(\vec{k},\eta)=\frac{iH}{\sqrt{2c_sP_{,X}k^3}}(1+ikc_s\eta)e^{-ikc_s\eta}
\label{modes}
\ee
Then from Eq.(\ref{keldysh}), and taking $\eta\rightarrow0$,
\ba
\Expect{\pi(\vec{k}_1)\pi(\vec{k}_2)\pi(\vec{k}_3)}&=&(2\pi)^3\frac{\delta^3(\vec{k}_1+\vec{k}_2+\vec{k}_3)}{(k_1k_2k_3)K^3} \left(\frac{3f_1}{4M^4}\right)\left(\frac{H^5}{\dot{\phi}^2c_s^2P_{,X}^3}\right)\hat{B}(\vec{k}_1,\vec{k}_2,\vec{k}_3)\\\nonumber
\Expect{\pi(\vec{k}_1)\pi(\vec{k}_2)\pi(\vec{k}_3)\pi(\vec{k}_4)}&=&(2\pi)^3\frac{\delta^3(\vec{k}_1+\vec{k}_2+\vec{k}_3+\vec{k}_4)}{(k_1k_2k_3k_4)K^5}\left(\frac{f_1}{8M^4}\right)\left(\frac{H^8}{\dot{\phi}^4c_sP_{,X}^4}\right)\hat{Q}(\vec{k}_1,\vec{k}_2,\vec{k}_3,\vec{k}_4)\\\nonumber
\ea
where $K=\sum_i k_i$, and $\hat{B}(\vec{k}_1,\vec{k}_2,\vec{k}_3)$ and $\hat{Q}(\vec{k}_1,\vec{k}_2,\vec{k}_3,\vec{k}_4)$ contain terms of order 1 as well as terms suppressed by powers of $c_s^2$. At this order, we may use $\zeta=-H\pi$. We also use the relations $\mathcal{P}_{\zeta}=\frac{H^4}{2\pi\dot{\phi}_0^2c_sP_{,X}}$ and
\ba
P_{,X}\left(\frac{1}{c_s^2}-1\right)&=&2XP_{,XX}\\\nonumber
&=&\frac{16f_1}{M^4}
\label{Pxsub}
\ea
Then for a general action $P(X,\phi)$ we can write (absorbing numerical factors into the functions of momenta)
\ba
\Expect{\zeta(\vec{k}_1)\zeta(\vec{k}_2)\zeta(\vec{k}_3)}&=&(2\pi)^3\frac{\delta^3(\vec{k}_1+\vec{k}_2+\vec{k}_3)}{(k_1k_2k_3)K^3}\left(\frac{\mathcal{P}_{\zeta}^2}{c_s^2}\right)B(\vec{k}_1,\vec{k}_2,\vec{k}_3)\\\nonumber
\Expect{\zeta(\vec{k}_1)\zeta(\vec{k}_2)\zeta(\vec{k}_3)\zeta(\vec{k}_4)}&=&(2\pi)^3\frac{\delta^3(\vec{k}_1+\vec{k}_2+\vec{k}_3+\vec{k}_4)}{(k_1k_2k_3k_4)K^5}\left(\frac{\mathcal{P}_{\zeta}^3}{c_s^4}\right)Q(\vec{k}_1,\vec{k}_2,\vec{k}_3,\vec{k}_4)
\ea
where terms in $B$ and $Q$ are $\mathcal{O}(c_s^0,c_s^2,c_s^4\dots)$ and so the term with the most powers of $c_s$ in the denominator scales as we expected: $\Expect{\zeta^n}\propto\mathcal{P}_{\zeta}^{n-1}/(c_s^2)^{n-2}$.

\subsection{Statistics of the action expanded to six-derivative terms}
Expanding the action up to six derivatives introduces a 5-point and 6-point function and gives tree-level corrections to the three and four-point correlations calculated above. The solutions for the Hubble parameter $H(\phi)$ is very nearly unchanged since the kinetic terms do not dominate the energy density, but the relationship between the sound speed and the function $f_1$ does change. For terms other than the leading one, we can no longer use just the first order expression relating $\delta\phi$ and $\zeta$ or the approximation $\dot{\pi}=\dot{\delta\phi}/\dot{\phi}_0$. For that reason, we will only write the contribution from the dominant term. A more precise calculation of the four-point correlation for $P(X,\phi)$ models, which also covers the case where derivative self-interactions do not dominate, can be found in \cite{Arroja:2008ga}.

To find the additional contribution from $X^3/M^8$, one can replace the functions $f_i$ by combinations of the derivatives $P_{,X}$, $P_{,XX}$ etc. appropriate for each order of the expansion. 

The dominant contribution from the new term containing second derivatives is
\ba
\Expect{\zeta(\vec{k}_1)\zeta(\vec{k}_2)\zeta(\vec{k}_3)}_{new}&=&(2\pi)^3\frac{\delta^3(\vec{k}_1+\vec{k}_2+\vec{k}_3)}{(k_1k_2k_3)K^3}\tilde{B}(\vec{k}_1,\vec{k}_2,\vec{k}_3) \left(-\frac{H}{\dot{\phi}}\right)^3\left(\frac{12f_3H^7\dot{\phi}}{8M^6P_{,X}^3c_s^4}\right)\\\nonumber
&=&(2\pi)^3\frac{\delta^3(\vec{k}_1+\vec{k}_2+\vec{k}_3)}{(k_1k_2k_3)K^3}\tilde{B}(\vec{k}_1,\vec{k}_2,\vec{k}_3) \left(\frac{\mathcal{P}_{\zeta}^2}{c_s^2}\right)\left(\frac{H}{c_sM}\right)^2\left(\frac{24f_3}{f_1+3f_3(X/M^4)}\right)
\ea
where we have used the first line of Eq.(\ref{Pxsub}) and 
\be
P_{,XX}=\frac{8f_1}{M^4}+\frac{48Xf_3}{M^8}
\ee
For $f_1(\phi/M)$ and $f_3(\phi/M)$ order 1, this is the coefficient we anticipated in Eq.(\ref{thirdorder}).

\bibliographystyle{JHEP}

\providecommand{\href}[2]{#2}\begingroup\raggedright\endgroup

\end{document}